\documentclass[fleqn,12pt,twoside]{article}
\usepackage[square,comma,numbers,sort&compress]{natbib}
\usepackage{espcrc1}
\usepackage{epsf}

\usepackage{graphicx}
\usepackage[figuresright]{rotating}

\newcommand{\AmS}{{\protect\the\textfont2
  A\kern-.1667em\lower.5ex\hbox{M}\kern-.125emS}}
\def\la{\;\raise0.3ex\hbox{$<$\kern-0.75em\raise-1.1ex\hbox{$\sim$}}\;}
\def\ga{\;\raise0.3ex\hbox{$>$\kern-0.75em\raise-1.1ex\hbox{$\sim$}}\;}
\newcommand{\msun}{\mbox{$M_\odot$}}

\title{Neutron star cooling}

\author{D.G.\ Yakovlev\address[Ioffe]{Ioffe Physico-Technical Institute,
        Politekhnicheskaya 26,
        194021 St.\,Petersburg, Russia},
        O.Y.\ Gnedin\address[STSI]{Ohio State University,
	760 1/2 Park Street, Columbus, OH 43215, USA},
        M.E.\ Gusakov\addressmark[Ioffe],
	A.D.\ Kaminker\addressmark[Ioffe],
        K.P.\ Levenfish\addressmark[Ioffe],
	and
	A.Y.\ Potekhin\addressmark[Ioffe]}

\begin{document}

\maketitle

\begin{abstract}
The impact of nuclear physics theories on cooling
of isolated neutron stars is analyzed.
Physical properties of neutron star matter
important for cooling are reviewed such as
composition, the equation of state,
superfluidity of various baryon species,
neutrino emission mechanisms.
Theoretical results are compared with
observations of thermal radiation from neutron stars.
Current constraints on
theoretical models of dense matter,
derived from such a comparison, are formulated.
\end{abstract}

\section{INTRODUCTION}
\label{introduction}

Our knowledge of neutron star (NS)
interiors is currently uncertain. In particular,
the fundamental problem of the equation of
state (EOS) at supranuclear densities
in NS cores is still unsolved.
Microscopic theories of dense matter are model dependent
and give a large scatter of possible EOSs
(e.g., Ref.\ \cite{haensel03}), from stiff to soft
ones, with different compositions of
inner NS cores (nucleons, hyperons, pion or kaon
condensates, quarks).
Thermal evolution of NSs
depends on a model of dense matter
which enables one to constrain the fundamental
properties of dense matter by comparing
simulations of NS cooling with observations.

NSs are born hot in supernova explosions,
with the internal temperature $T \sim 10^{11}$ K.
In about one minute after the birth a star becomes
transparent for neutrinos generated
in its interiors. In the
following neutrino-transparent stage
the star cools via
neutrino emission from the entire stellar body
and via heat transport through the envelope to the surface and
subsequent thermal surface emission of photons.
In a hundred of years the NS crust and core
become thermally adjust and the NS interior becomes
isothermal (with the only temperature gradient located near the surface).
After that the effective surface temperature, $T_s$,
reflects the thermal state of the core.

The recent development of the
theory has been reviewed,
e.g., in Refs.\ \cite{yp04,pageetal04}.
Here, we outline the current status of the problem.

\section{OBSERVATIONS}
\label{observations}

\begin{figure}[t]
\begin{minipage}[t]{77mm}
\epsfysize=70mm
\epsffile[20 150 570 690]{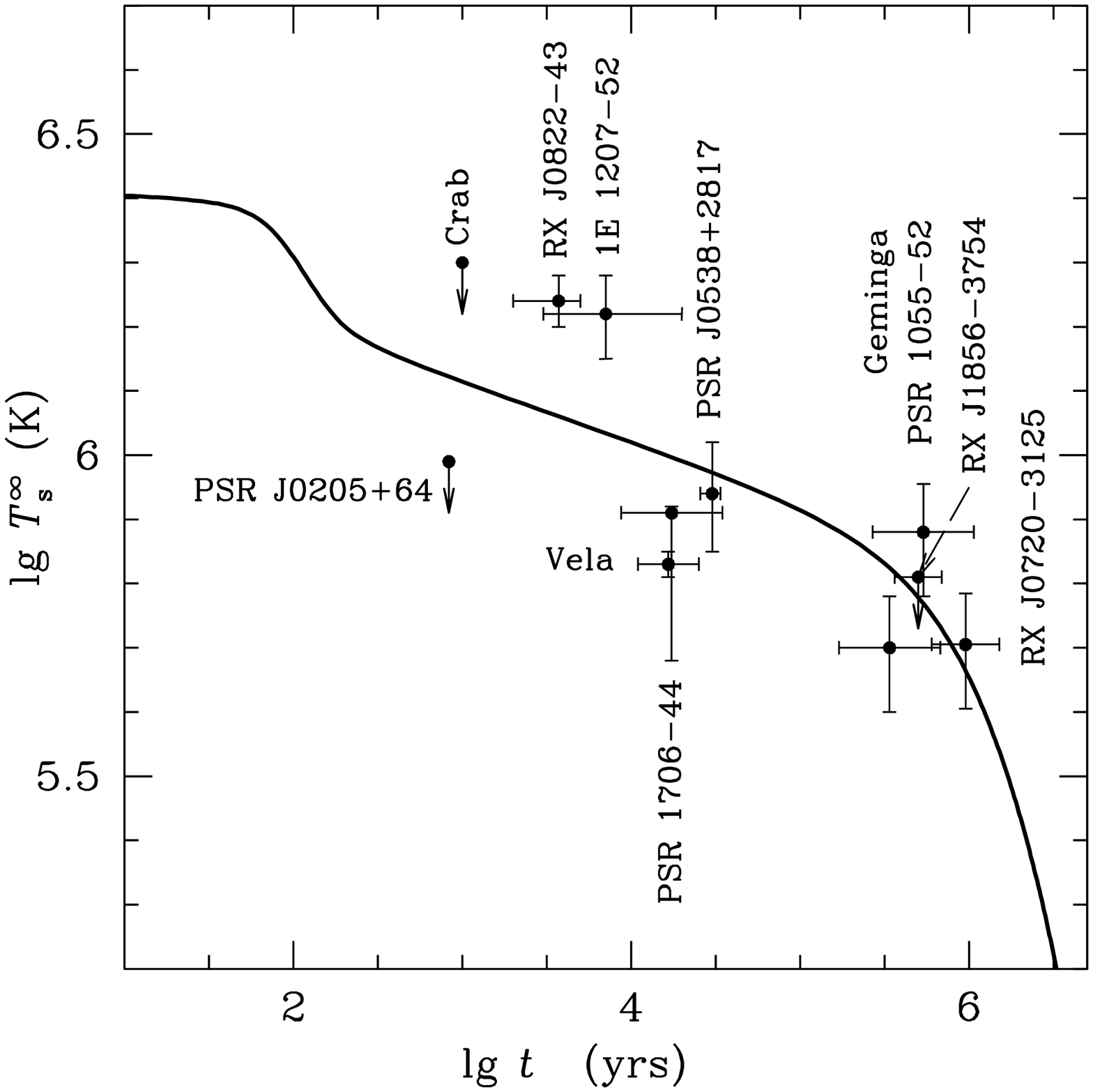}
\caption{
Observational limits of surface temperatures for several
isolated NSs compared with the basic theoretical
cooling curve of a non-superfluid NS model.
}
\label{obs}
\end{minipage}
\hspace{\fill}
\begin{minipage}[t]{77mm}
\epsfxsize=70mm
\epsffile[20  150  560 690]{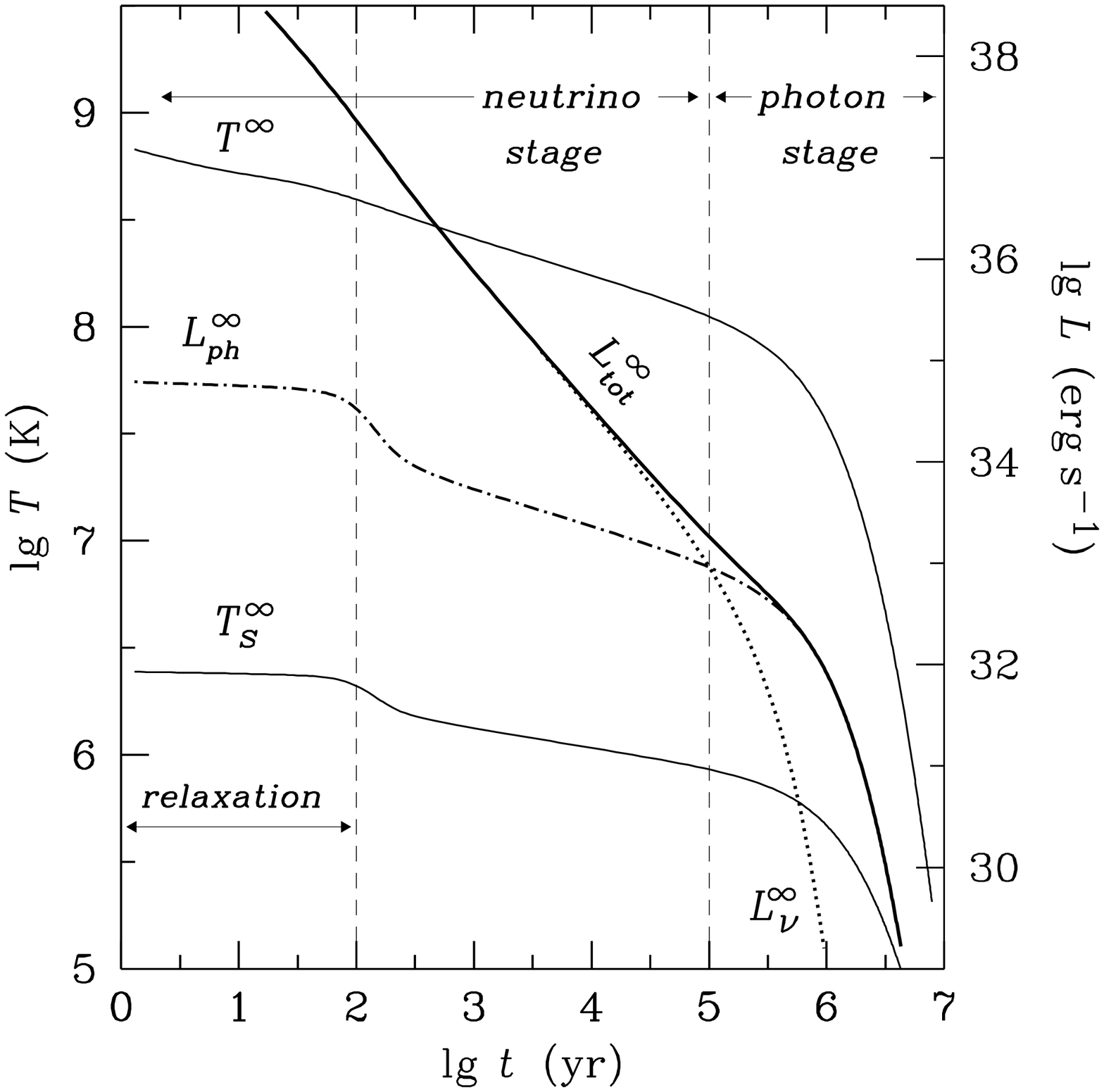}
\caption{Internal and surface temperatures;
neutrino, photon and total luminosities
(redishifted for a distant observer) for the same
NS model as in Fig.\ \ref{obs}.
}
\label{coola}
\end{minipage}
\end{figure}

Observations of isolated NSs,
whose thermal surface radiation has been detected or constrained,
are summarized in Fig.\  \ref{obs} (following  Ref.\ \cite{gusakovetal04}).
We present the estimated NS ages $t$ and
effective surface temperatures $T_s^\infty$
(as detected by a distant observer).

For the two youngest sources
only
upper limits on
the surface temperature $T_s^\infty$ have been established
\cite{slane04,weisskopf04}.
The surface temperatures of the next five sources,
with ages $10^3 \la t \la 10^5$ years,
have been obtained
\cite{zavlinetal99,zavlinetal04,pavlovetal01,mcgowanetal04,zp04}
by fitting their thermal radiation spectra with
hydrogen atmosphere models.
Such models are more consistent with other information
on these sources (e.g., Ref.\ \cite{pz03})
than the blackbody model.
On the contrary, for Geminga and PSR B1055--52
we present the values
of $T_s^\infty$  \cite{zp04,pz03} inferred using the blackbody
spectrum because this spectrum is more consistent for these sources.
The surface temperature of RX J1856.4--3754 is still
uncertain.
Following \cite{gusakovetal04} 
we adopt the upper limit $T_s^\infty
< 0.65$ MK.  Finally, $T_s^\infty$ for RX J0720.4--3125
is taken from
Ref.\ \cite{motchetal03},
where the observed spectrum is interpreted
with a model of a hydrogen atmosphere of finite depth.

As seen from Fig.\ \ref{obs}, observational limits scatter
in the $T_s^\infty-t$ plane.
What can be learnt on
dense matter in NS interiors from this scatter?

\section{THEORY VERSUS OBSERVATIONS}
\label{basic}

A neutron star consists of a thin crust
(of mass $\la 10^{-2} \msun$, where $\msun$ is the solar mass)
and a core (e.g., Ref.\ \cite{haensel03}).
The core-crust interface is placed at the mass density
$\rho \sim \rho_0/2$, where $\rho_0\approx 2.8 \times 10^{14}$
g~cm$^{-3}$ is the density of saturated nuclear matter.
The crustal matter contains atomic nuclei, electrons, and
(at $\rho \ga 4 \times 10^{11}$ g~cm$^{-3}$) free neutrons.
The core is further subdivided into the outer ($\rho \la 2 \rho_0$)
and inner parts. The outer core consists of neutrons, with an
admixture of protons, electrons, and muons.
The composition of the inner core is still unknown. It
may be the same composition as the outer core but
may also contain hyperons,
pion or kaon condensates,
quark matter, or a mixture of different phases.

NS matter is strongly degenerate. The EOS, NS
masses $M$ and radii are almost temperature independent.
Low-mass NSs ($M \sim \msun$) have rather low
central densities $\rho_{\rm c} \la 2 \, \rho_0$ and do not
possess inner cores. NSs with masses close to the maximum
allowable mass ($M_{\rm max} \sim (1.5-2.5)\,\msun$, for different
model EOSs) have massive inner cores.

NS cooling is calculated with a cooling code (e.g.,
Ref.\ \cite{gyp01}) in the form of {\it cooling curves},
$T_s^\infty(t)$
(e.g., Fig.\ \ref{obs}).
The initial cooling stage, $t \la 100$ years, is accompanied
by thermal relaxation of NS interiors
(Fig.\ \ref{coola}). As long as $t \la 10^5$ years,
a star cools mainly via neutrino emission from its interiors
(mainly from the core); this is the {\it neutrino cooling stage}.
Later, at $t \ga 10^5$ years, the neutrino emission becomes
inefficient, and the star cools via thermal surface emission
of photons (the {\it photon cooling stage}).

NSs may have different masses,
surface magnetic fields, composition of surface layers,  etc.,
but they are supposed to have the same EOS and superfluid
properties of internal layers.
In the absence of exact
microscopic theory of NS matter we will
use several model EOSs and phenomenological superfluidity models.
\begin{figure}[t]
\epsfysize=70mm
\epsffile[40 180 570 470]{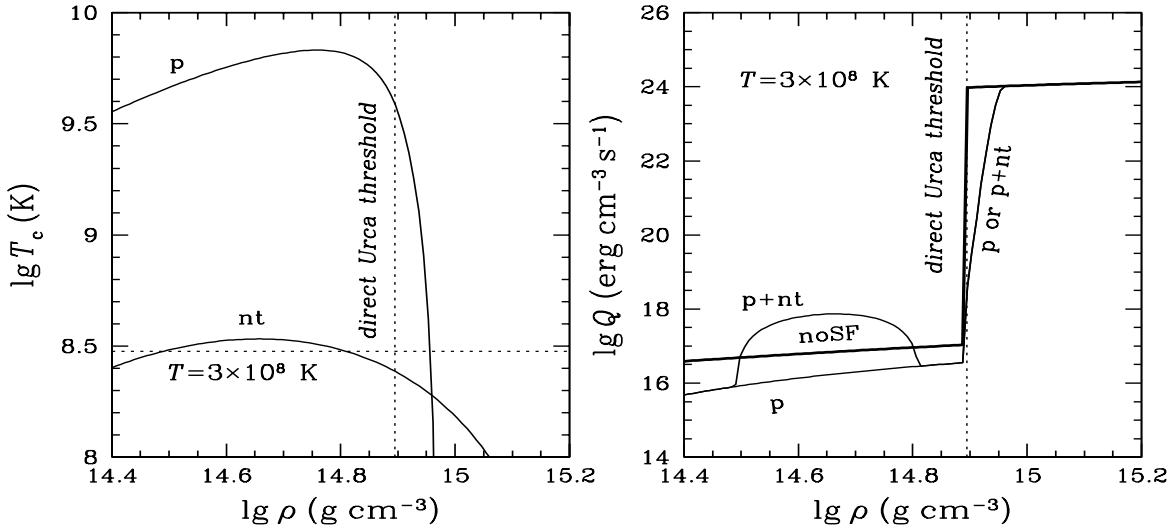}
\caption{
{\em Left:} Illustrative models of critical temperatures
for proton (p) and neutron (nt) pairing in NS core.
{\em Right:} Neutrino emissivity in the same NS core
at the temperature $T=3 \times 10^8$ K for non-superfluid matter
(thick line; noSF) and in the presence of either proton pairing (p)
or proton and neutron pairing (p+nt). The vertical dotted line
indicates the threshold of the direct Urca process.
}
\label{supnu}
\end{figure}

The main regulators of NS cooling are:\\
({\it a}) EOS and composition
of NS cores which affect neutrino emission mechanisms;\\
({\it b}) Superfluidity of baryons in NSs --- it affects neutrino
emission and heat content;\\
({\it c}) The presence of light elements
(accreted envelopes) and strong magnetic
fields in NS surface layers. These factors affect the thermal conductivity
and the relation between the internal and surface
temperatures of the star.\\
We discuss these regulators below (except for magnetic fields
whose effects are examined in \cite{pycg03}). Other regulators are reviewed,
e.g., in Ref.\ \cite{yp04}.

Figure \ref{obs}
shows
the
{\it basic cooling curve}. It is calculated for a star
with a non-superfluid nucleon core, where
the powerful direct Urca process of neutrino emission is forbidden.
Such a star cools mainly via neutrino emission
produced by the less powerful modified Urca
process; the accretion envelope is absent.
This {\it basic curve} is {\it universal},
being almost independent of the EOS and NS mass.
It cannot explain
all
the observations -- some
NSs are hotter and some colder than predicted
by the curve.
However, one
can explain the data
by employing other cooling regulators.

The effects of
superfluidity and the direct Urca process,
are demonstrated
in Fig.\ \ref{supnu}.
Here we adopt a
moderately
stiff EOS of dense nucleon matter suggested in Ref.\ \cite{pal88} (the same 
version
as used in \cite{yp04}).
This EOS
opens the
direct Urca process at
$\rho > \rho_{\rm D}=7.851 \times 10^{14}$ g~cm$^{-3}$, i.e.,
at $M> M_{\rm D}=1.358\,\msun$ ($M$ being
the gravitational mass)
and
gives NS models
with $M_{\rm max}=1.977\,\msun$.
In non-superfluid matter
the direct Urca process switches on sharply at $\rho > \rho_{\rm D}$.
However, neutrons and protons (like other baryons) in NS cores can
be in superfluid state. As a rule, neutrons undergo
triplet-state pairing, whereas protons undergo singlet-state
pairing
(e.g., Ref.\ \cite{ls01})
with density dependent critical
temperatures $T_{\rm cnt}(\rho)$ and
$T_{\rm cp}(\rho)$ which are extremely sensitive to
theoretical models. Superfluidity
suppresses traditional neutrino emission mechanisms
(the modified and direct Urca processes and nucleon-nucleon bremsstrahlung)
but opens a new neutrino process associated
with Cooper pairing of baryons \cite{frs76}.
Figure \ref{supnu} shows some phenomenological $T_{\rm cnt}(\rho)$ and
$T_{\rm cp}(\rho)$ curves (from Ref.\ \cite{yp04}) and demonstrates that
the direct Urca process and superfluidity greatly affect
the neutrino emission (and, hence, NS cooling, as discussed later).

The cooling can be strongly different for low mass,
medium mass, and high mass NSs.

\subsection{Cooling of low-mass stars}
\label{low-mass}

\begin{figure}[htb]
\begin{minipage}[t]{77mm}
\epsfxsize=73.4mm
\epsffile[20 130 570 690]{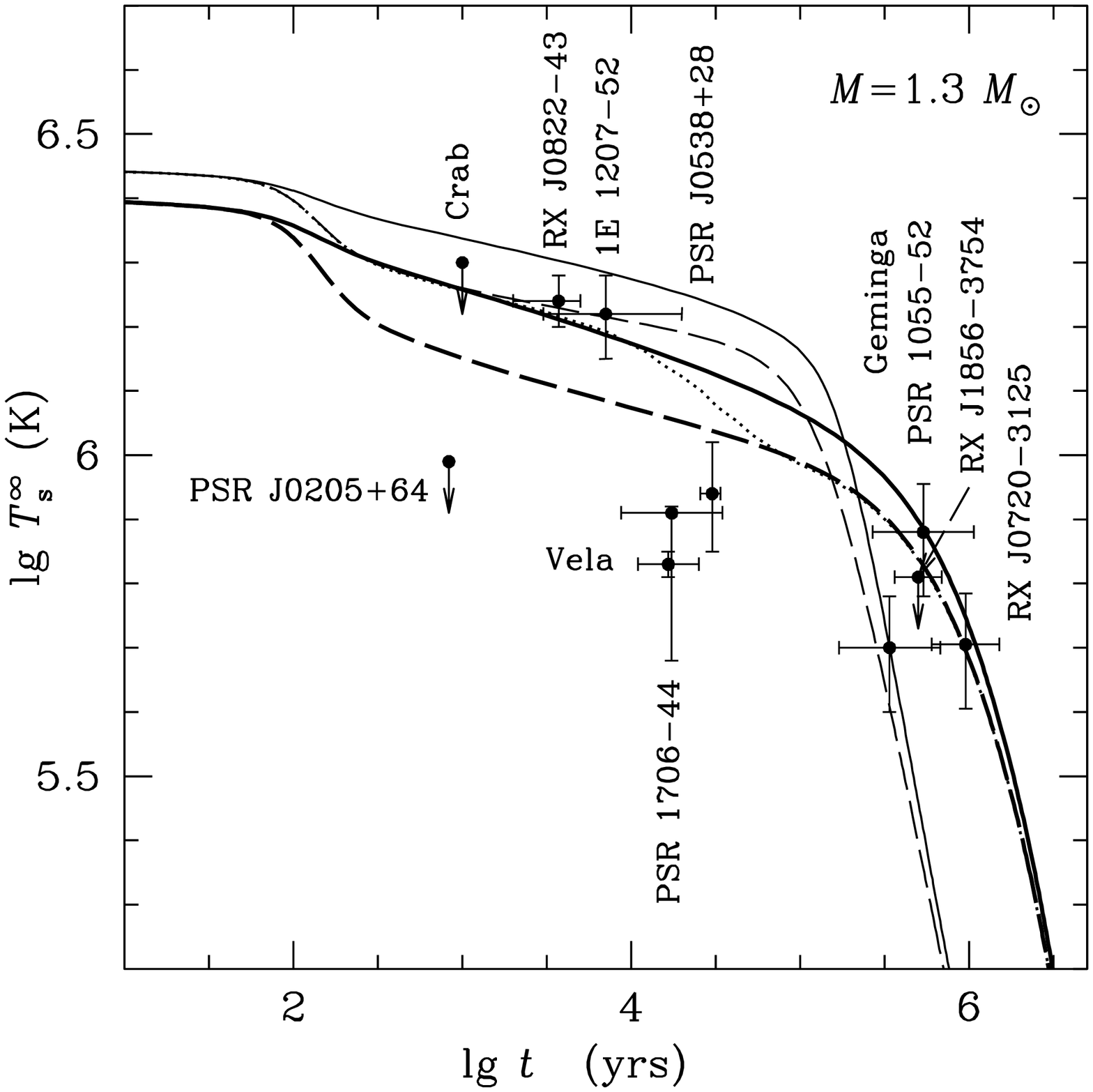}
\caption{Cooling curves for the star with $M=1.3\, \msun$ and
nucleon core compared with the observations
(see text for other explanations).
}
\label{fig-low-mass}
\end{minipage}
\hspace{\fill}
\begin{minipage}[t]{77mm}
\epsfxsize=77mm
\epsffile[70 205 515 640]{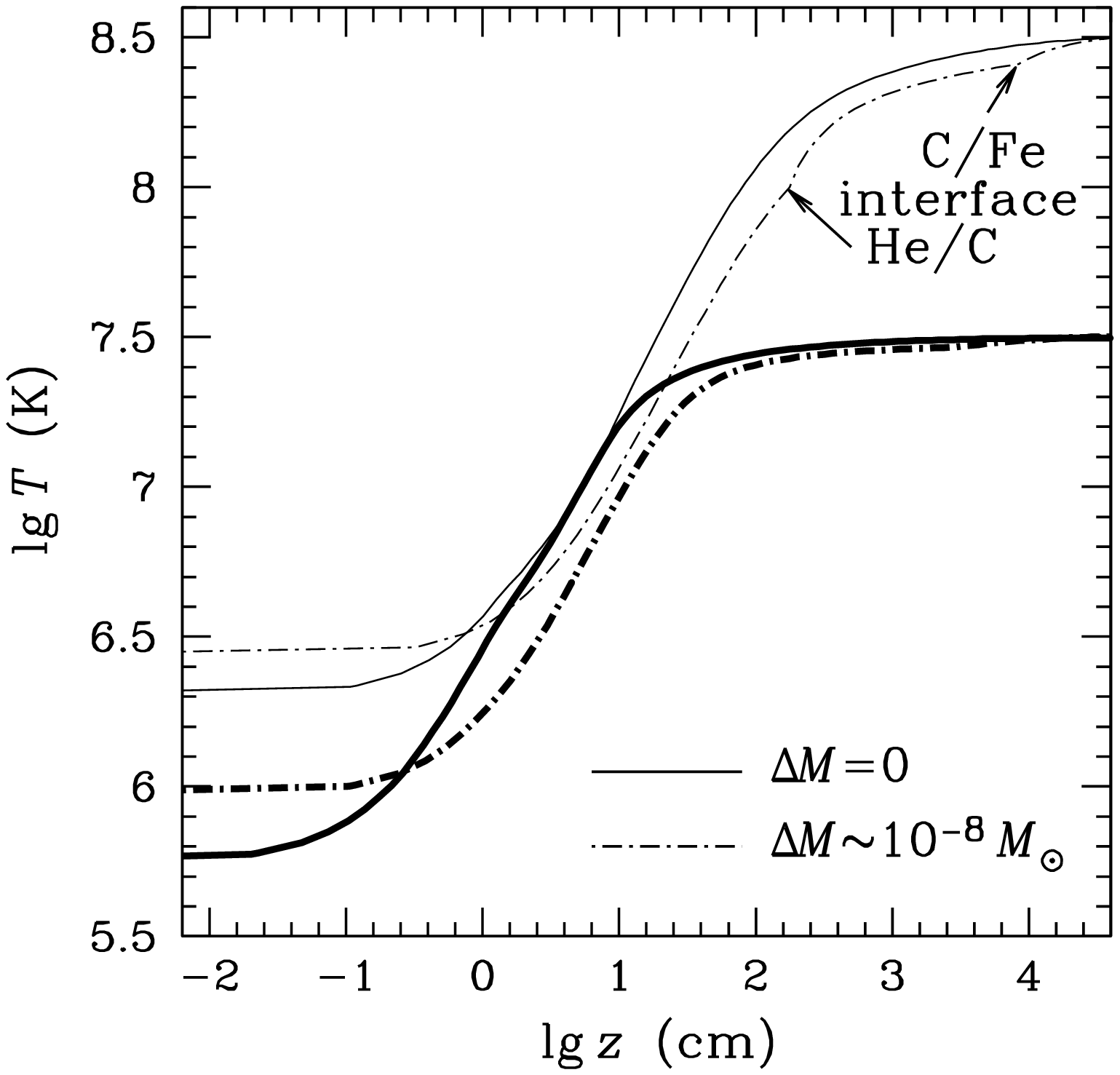}
\caption{Temperature versus depth $z$ in the
NS envelope
for two internal temperatures: $10^{8.5}$~K
(light curves) and $10^{7.5}$~K (heavy curves), for
nonaccreted and accreted envelopes.
}
\label{fig-prf}
\end{minipage}
\end{figure}

Low-mass NSs possess only
outer nucleon cores. Some cooling curves are presented
in Fig.\ \ref{fig-low-mass} for NS models constructed with the same EOS as
in Fig.\ \ref{supnu}.

The solid curves are calculated assuming
strong proton superfluidity p. It
suppresses the modified Urca process in a low-mass NS.
The neutrino luminosity of the star becomes
lower (Fig.\ \ref{supnu}), being determined by
a weaker mechanism of neutrino emission
(neutron-neutron bremsstrahlung,
unaffected by superfluidity as long as neutrons
are non-superfluid). This rises the cooling curves at the
neutrino cooling stage.
The thick solid curve is calculated
for a star without any accreted envelope. This curve
(contrary to the basic curve in Fig.\ \ref{obs})
goes high enough to explain the sources hottest for their
age (RX J0822--43, 1E 1207--52, PSR B1055--52).
Thus, we may treat these sources as low-mass NSs.
The thin solid curve is calculated assuming, additionally,
the presence of
hydrogen or helium accreted envelope
of mass $\Delta M =10^{-8}\,\msun$.
Light elements increase the thermal conductivity
of the surface layers which further rises $T_s^\infty$
at the neutrino cooling stage. The thin solid curve
is close to the highest cooling curve provided by the standard
cooling theory of NSs.

\begin{figure}[t]
\epsfysize=70mm
\epsffile[20 150 465 415]{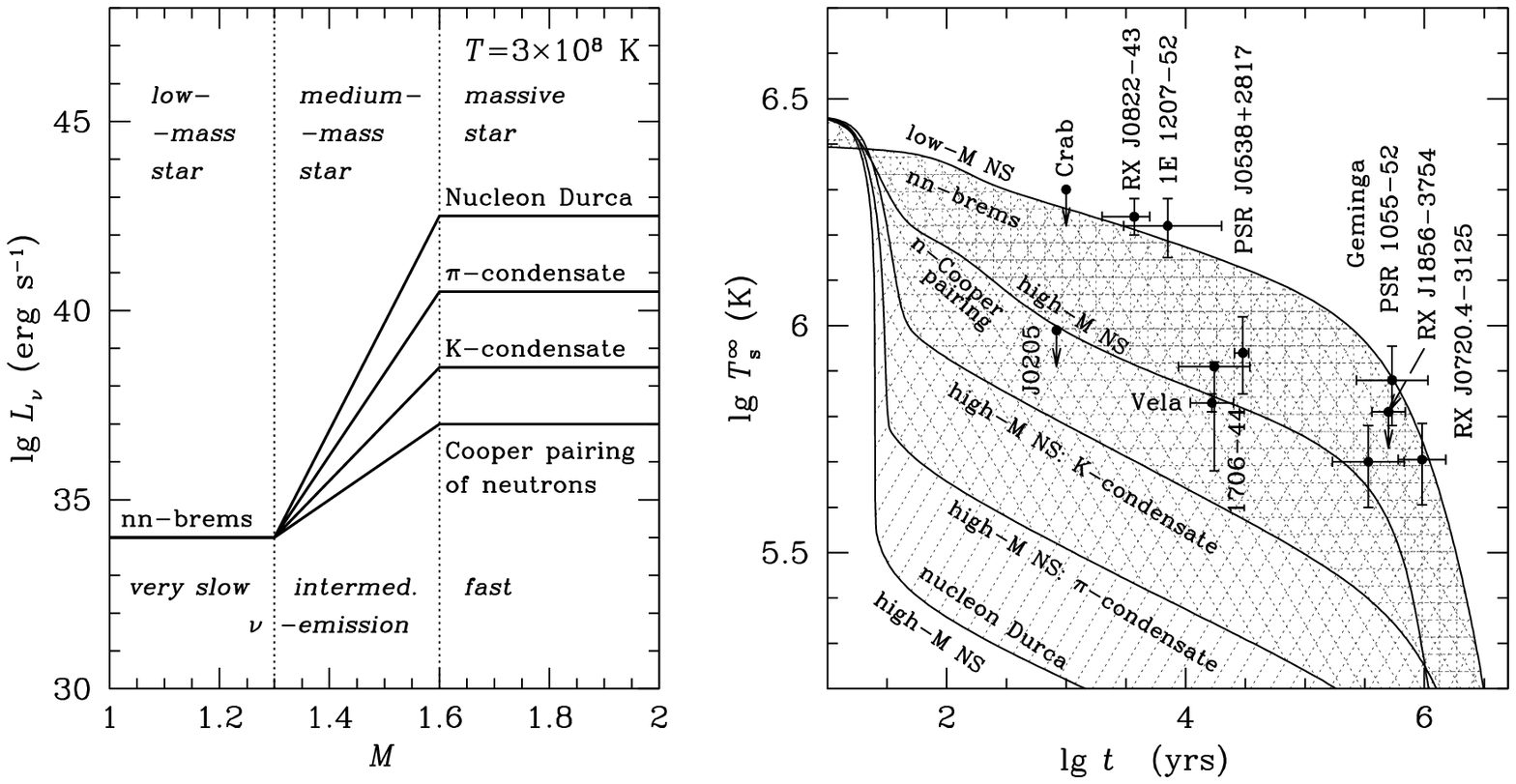}
\caption{ {\em Left:} A sketch of the neutrino luminosity
$L_\nu$ versus stellar mass for NSs with the internal
temperature $T=3 \times 10^8$ K at four models of
NS structure. {\em Right:} Four hatched regions
of $T_s^\infty$ which can be explained by cooling
of NSs of different masses for four models in the left panel.
}
\label{four}
\end{figure}

Actually, we do not need very strong proton superfluidity 
(such as model p) to interpret the observations.
The dashed lines in Fig.\ \ref{fig-low-mass} are the same as the solid 
lines,
but the proton critical temperature $T_{\rm cp}(\rho)$
is reduced by a factor of 6. Weaker proton superfluidity
relaxes the superfluid suppression of the modified Urca processes
and lowers the cooling curves (with respect to the solid ones).
Nevertheless, such superfluidity is still sufficient to interpret
the observations:
old hot sources are consistent with the thick dashed line 
(no accretion envelope) while young hot sources 
are well explained assuming an accreted envelope (thin dashed line).
Finally, the dotted curve is the same as the thin dashed curve
but the mass of the accreted envelope is assumed to decrease
with time  (e.g., due to diffuse burning \cite{cb03})
as $\Delta M(t)=\Delta M(0)\,\exp(-t/\tau)$,
with $\tau=4000$ years.
This model can also
explain the observations of NSs hottest for their ages.

The effect of light-element envelopes on the cooling
can be understood from Fig.~\ref{fig-prf}. It shows
the growth of temperature within the NS of mass $M=1.4\,\msun$
and radius $R=10$ km. 
The solid lines refer to
a non-accreted (Fe) surface. The dot-and-dashed lines are for
the thickest accreted (He and C) envelope that can survive
with respect to nuclear burning.
Because He and C have higher thermal conductivity than
Fe, their presence makes the NS envelope more
heat-transparent, increasing $T_s$.

\begin{figure}[t]
\epsfysize=70mm
\epsffile[30 160 520 400]{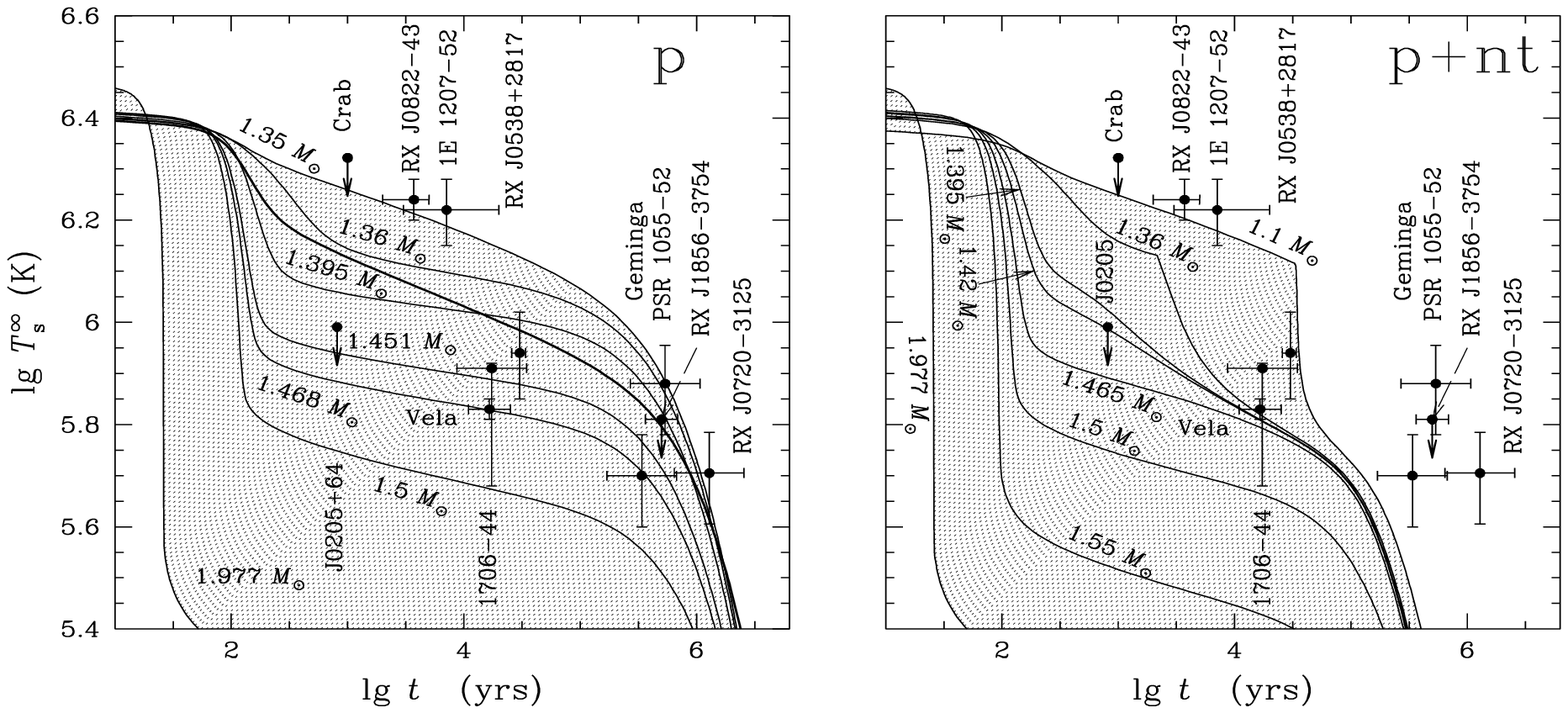}
\caption{
{\em Left:} Cooling of NSs of several masses
(indicated near the curves). NSs are assumed to have
nucleon cores and proton superfluidity p from Fig.\ \ref{supnu}.
{\em Right:} Same as in the left panel but adding the
effect of neutron superfluidity nt.
}
\label{fig-weigh}
\end{figure}

\subsection{Cooling of high-mass neutron stars}
\label{high-mass}

\begin{figure}[t]
\epsfysize=70mm
\epsffile[20 150 575 420]{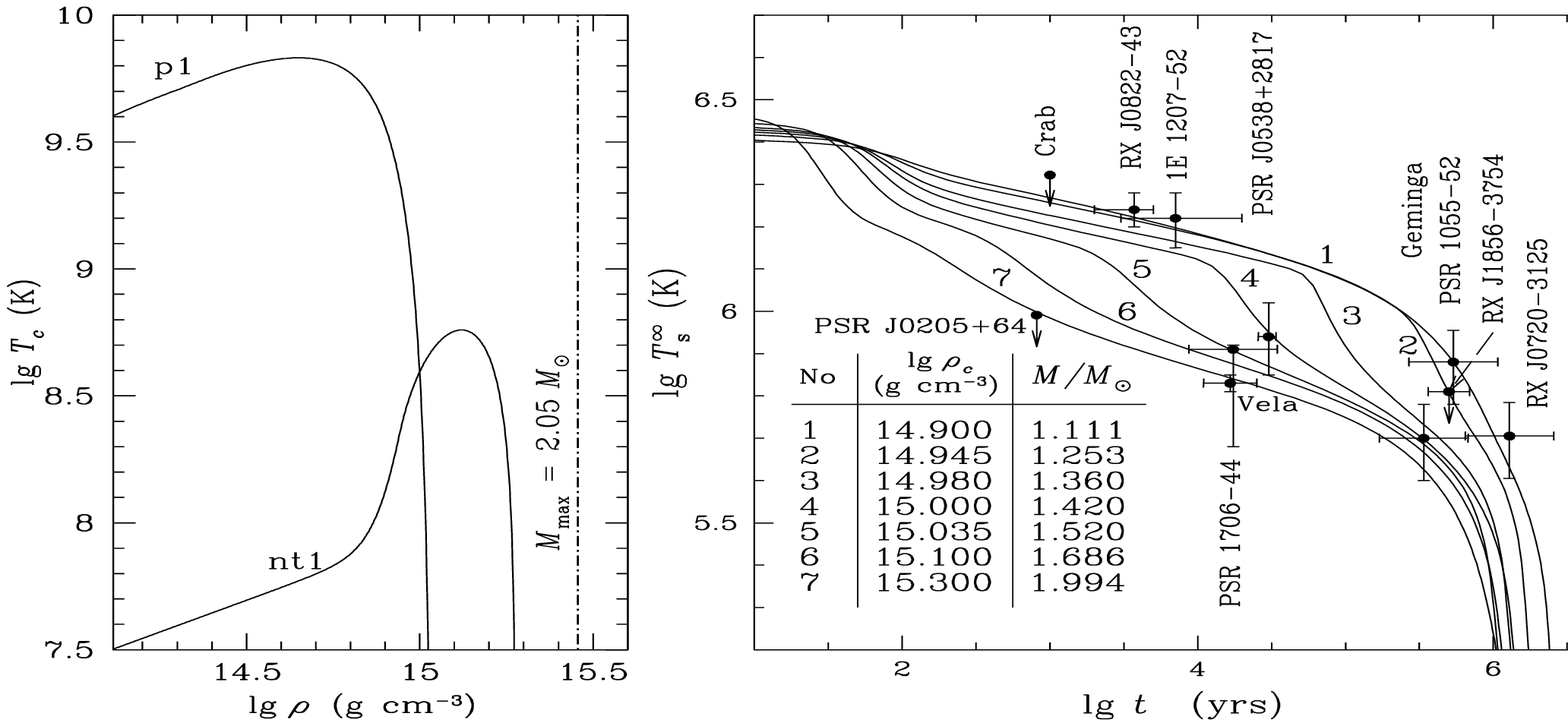}
\caption{
{\em Left:} Model density dependence of critical temperatures
of protons (p1) and neutrons (nt1) in a nucleon NS core for the EOS
which forbids the direct Urca process.
{\em Right:} Cooling curves of NSs of several masses
for the same EOS, taking into account superfluidities p1 and nt1.
After Ref.\ \cite{gusakovetal04}.
}
\label{fig-cooper}
\end{figure}

High-mass NSs have large central densities,
masses $M \sim M_{\rm max}$, and contain
massive inner cores. Microscopic theories predict that, as a rule,
superfluidity dies out (and does not suppress
neutrino emission) in the central parts of
such stars.

One can propose very different cooling scenarios of
high-mass NSs. The simplest scenario assumes non-superfluid
nucleon cores where the direct Urca process
is forbidden. The corresponding cooling curves would
be the same as the basic curve in Fig.\ \ref{obs};
they cannot explain all the observations.

It is widely thought that the neutrino emission
in high-mass NSs is {\it enhanced} as compared to the emission
provided by the modified Urca process. An enhanced emission
would lead to {\it fast} cooling, allowing one to explain
the observations of NSs coldest for their ages.
There are different enhancement levels
for different models of NS internal structure.
Four scenarios are presented in Fig.\ \ref{four}.
The left panel is a rough sketch of the neutrino luminosity
as a function of $M$ at $T=3 \times 10^8$ K. One can generally assume
a slow neutrino emission in low-mass NSs
(e.g., provided by the neutrino bremsstrahlung in neutron-neutron
collisions), an enhanced neutrino emission in high-mass NSs,
and the transition from the slow to enhanced emission
with increasing $M$ in medium-mass NSs. The mass range
of medium-mass stars is model dependent \cite{kyg02}.
In massive NSs $L_\nu$ scales as $T^6$ for all scenarios except
for Cooper pairing scenario
(where $L_\nu \propto T^8$ \cite{gusakovetal04,pageetal04}).
In low-mass NSs, $L_\nu \propto T^8$.

The right panel of Fig.\ \ref{four} shows the limiting cooling curves for
each scenario (no accreted envelopes). The upper curve
refers to low-mass NSs. Their cooling is the same for all scenarios,
and it explains the observations of NSs hottest for their age
(Sect.\ \ref{low-mass}).
Four lower curves show cooling of maximum-mass NSs for four scenarios.
They are the lowest cooling curves in these scenarios.
The three lowest curves are taken from Ref.\ \cite{yp04},
while the fourth is from  Ref.\ \cite{gusakovetal04}.
The range of $T_s^\infty$ between the upper curve and a
lower curve can be filled by cooling curves of NSs with masses
from $\sim \msun$ to $M_{\rm max}$.
Thus, we have four different ranges of $T_s^\infty$ (four hatched
regions) for four scenarios.

The highest enhancement of neutrino emission is provided
by the direct Urca (Durca) process  in nucleon (or nucleon-hyperon) cores
\cite{lpph91}.
This scenario predicts the coldest massive NSs and the widest theoretical
$T_s^\infty$ range. If the direct Urca process is forbidden
but pion condensate is present in the inner NS core, the
enhancement of neutrino emission is provided by
the process of Durca type involving quasi-nucleons
(e.g., \cite{pethick92}).
This enhancement is weaker, the massive stars
are hotter, and the acceptable $T_s^\infty$ range narrower.
If pion condensate is absent, but kaon condensate available,
the neutrino emission enhancement
(in Durca-type processes involving quasi-baryons \cite{pethick92})
is even weaker
and the $T_s^\infty$ range narrower. Nearly the same enhancement
is expected in NSs with non-superfluid inner cores composed
of quarks. Finally, the lowest enhancement can be produced in
nucleon inner cores
\cite{pageetal04,gusakovetal04},
where the direct Urca process is forbidden
but mild superfluidity (e.g., of neutrons) is available
(see Sect.\ \ref{sect-cooper}).
It triggers the Cooper pairing neutrino emission
which accelerates NS cooling.
It gives the narrowest
theoretical region of $T_s^\infty$.

As seen from Fig.\ \ref{four}, all four scenarios are compatible with
the observations.

\subsection{Cooling of medium-mass neutron stars}
\label{medium-mass}

The next question, crucial
for explaining the observations
(e.g., of the Vela and Geminga pulsars),
is how cooling curves fill
hatched regions in Fig.\ \ref{four} if we vary the NS mass from $\sim \msun$
to $M_{\rm max}$. The answer \cite{yh03}
is closely related to the
contrast of slow and enhanced neutrino luminosities
and the mass range of medium-mass NSs in Fig.\ \ref{four}.
Let us outline this problem for NSs with nucleon cores.

Figure \ref{fig-weigh} shows cooling of NSs of several masses
with the same EOS as in Fig.\ \ref{supnu} (no accretion envelopes).
In the left panel we take into account strong proton pairing p,
which extends to densities $\rho > \rho_{\rm D}$.
As long as  $T_{\rm cp}(\rho) \ga 3 \times 10^9$ K,
it suppresses the modified and even the direct Urca process
and leads to very slow neutrino emission.
At higher $\rho$ it gradually dies out, opening
the direct Urca process.
The gradual opening broadens the direct Urca threshold
(Fig.\ \ref{supnu}) and ensures the gradual decrease of cooling curves with
increasing $M$. In this way we may attribute masses
to observed NSs \cite{khy01}.
For instance,
we obtain $M \approx 1.47\,\msun$ for the Vela pulsar. However, this
{\it weighing} of NSs is sensitive
to the EOS of dense matter, the threshold
of the direct Urca process, and the superfluidity model $T_{\rm cp}(\rho)$.
Were superfluidity absent, the transition from slow
to fast cooling would occur in a very narrow mass interval
($0<M-M_{\rm D} \la 0.001\,\msun$), and
the successful interpretation of the data
would be unlikely (e.g., Ref.\ \cite{yp04}).

\subsection{Harmful and useful Cooper-pairing neutrino emission}
\label{sect-cooper}

The cooling effect of neutrino emission due to Cooper pairing
of nucleons may be different.
For instance, let us assume the presence of neutron
pairing nt with the peak of $T_{\rm cnt}(\rho)$ as low as
$\sim 4 \times 10^8$ K at $\rho \sim 4 \times 10^{14}$
g~cm$^{-3}$ (Fig.\ \ref{supnu}). This superfluidity is mild and insignificant,
according to nuclear physics standards, but crucial for
NS cooling. It appears in a cooling star when the internal
temperature falls below the peak value. It creates then a powerful
neutrino emission owing to Cooper pairing of neutrons
in outer NS cores (especially efficient in low-mass NSs).
The emission accelerates NS cooling, as shown in the right
panel of Fig.\ \ref{fig-weigh}, and violates the interpretation of the
observations of such sources as PSR B1055--52. Thus,
this mild neutron superfluidity contradicts the observations.

The opposite example is given in Fig.\ \ref{fig-cooper}.
Let us consider NSs with nucleon cores and employ the EOS
\cite{dh02}
which forbids the direct Urca
process in all NSs with $M \leq M_{\rm max}=2.05\,\msun$.
Furthermore, let us adopt the model of strong proton pairing p1
and mild neutron pairing nt1 (the left panel of Fig.\ \ref{fig-cooper}).
Pairing p1 is similar to pairing p in Fig.\ \ref{supnu};
it suppresses the modified Urca process in low-mass NSs.
The peak of $T_{\rm cnt}(\rho)$ for pairing nt1 is as low
as for pairing nt but shifted to higher densities.
Accordingly, pairing nt1 is inefficient in low-mass stars
and does not speed up their cooling. However, the enhanced
neutrino emission owing to this pairing
operates in massive NSs and accelerates their cooling
(a scenario considered in Sect.\ \ref{high-mass}, Fig.\ \ref{four}). Then,
as seen in the right panel of Fig.\ \ref{fig-cooper},
the cooling of NSs
of different masses enables us to explain the data.
In this case mild neutron pairing is useful for interpretation
of the observations but a successful interpretation
is possible only under stringent constraints
on the $T_{\rm cnt}(\rho)$ profile \cite{gusakovetal04}.
Moreover, a discovery of a new NS slightly colder than those observed now
would ruin this interpretation.

\section{CONCLUSIONS}
\label{conclusions}

We have outlined several possible scenarios of
NS cooling. In particular, we have considered cooling
of NSs with nucleon (nucleon/hyperon) cores, and
cores containing exotic phases
of dense matter.
We have shown that many
scenarios are currently compatible with
observations of thermal radiation from isolated NSs. Our main conclusions 
are:

(i) Some NSs (e.g., RX J0822--43 or PSR B1055--52)
are hotter and some (e.g., the Vela pulsar)
colder than non-superfluid NSs which cool via the modified
Urca process. Hotter NSs
are possibly low-mass stars, while
coldest observed NSs are possibly more massive.


(ii) Currently, the observations unable one to discriminate
between many cooling scenarios. However, they seem to rule out
mild superfluidity with the peak of the
critical temperature $T_{\rm c}(\rho)$ between
$\sim 3 \times 10^8$ and $\sim 2 \times 10^9$ K
at $\rho \la 8 \times 10^{14}$ g~cm$^{-3}$ in NS cores.
This superfluidity would initiate Cooper pairing neutrino
emission in low-mass NSs hampering the interpretation of
the observations of old and warm NSs, such as PSR B1055--52.
In contrast, mild superfluidity with the peak of $T_{\rm c}(\rho)$
at {\it higher} $\rho$ can be useful for interpretation of the observations.

We have discussed main cooling scenarios but not all
of them. Some others are reviewed in
Refs.\ \cite{yp04,pageetal04}.
Actually, the effects of superfluidity are more sophisticated
than discussed above. For instance, cooling curves do not
change
qualitatively
by exchanging $T_{\rm c}(\rho)$ for neutrons
and protons \cite{gus04}. 
Strong superfluidity of all baryon species
(with peaks of $T_{\rm c}(\rho)$ higher than $2\times 10^9$~K)
leads to a very
low heat capacity of NSs. Such stars appear at the photon cooling
stage earlier than at $t \sim 10^5$ years; they are too
cold at that stage.
Weak superfluidity, with $T_{\rm c}(\rho)
\la 3 \times 10^8$ K, does not occur in
NSs of ages $t \la 10^5$ years and does not affect
their cooling.
Cooling of low-mass NSs can be strongly affected by
singlet-state pairing of neutrons in inner stellar crusts.

New observations of NSs are required for a better understanding
of their internal structure. New discoveries of cold NSs would
be especially useful. Observations of cooling NSs
can be analyzed together with other observational data,
for instance, with observations of quiescent thermal emission from
NSs in soft X-ray transients (see \cite{yp04}, for references).
This would allow one to
obtain more stringent constraints on NS structure.


DY is grateful to C.\ Pethick for discussions and to
NORDITA for financial support which made his participation in INPC2004
possible. KL acknowledges the support of the Russian Science Support Foundation.
The work was partly supported by RFBR (grants 02-02-17668
and 03-07-90200), RLSSF (grant 1115.2003.2), and INTAS
(grant YSF 03-55-2397).


\begin{thebibliography}{9}

\bibitem{haensel03}
P.\ Haensel,
In: Final Stages of Stellar Evolution,
C.\ Motch and J.-M.\ Hameury (eds.),
EAS Publications Series: EDP
Sciences (2003) 249.


\bibitem{yp04}
D.G.\ Yakovlev and C.J.\ Pethick,
Ann.\ Rev.\ Astron.\ Astrophys.\ 42 (2004) 169.

\bibitem{pageetal04}
D.\ Page, J.M.\ Lattimer, M.\ Prakash, and A.W.\ Steiner ,
Astroph.\ J.\ (2004) submitted [astro-ph/0403657].

\bibitem{gusakovetal04}
M.E.\ Gusakov, A.D.\ Kaminker, D.G.\ Yakovlev, and O.Y.\ Gnedin,
Astron.\ Astrophys.\ 423 (2004) 1063.

\bibitem{slane04}
P.\ Slane, D.J.\ Helfand, E.\ van der Swaluw, and S.S.\ Murray,
Astrophys.\ J.\ (2004) submitted [astro-ph/0405380].

\bibitem{weisskopf04}
M.C.\ Weisskopf, S.L.\ O'Dell, F.\ Paerels, R.F.\ Elsner,
W.\ Becker, A.F.\ Tennant, and D.A. Swartz,
Astrophys.\ J.\ 601 (2004) 1050.

\bibitem{zavlinetal99}
V.E.\ Zavlin, J.\ Tr\"umper, and G.G.\ Pavlov,
Astrophys.\ J.\  525 (1999) 959.

\bibitem{zavlinetal04}
V.E.\ Zavlin, G.G.\ Pavlov, and D.\ Sanwal,
Astrophys.\ J.\ 606 (2004) 444.

\bibitem{pavlovetal01}
G.G.\ Pavlov, V.E.\ Zavlin, D.\ Sanwal,
V.\ Burwitz, and G.P.\ Garmire,
Astrophys.\ J.\  552 (2001) L129.

\bibitem{mcgowanetal04}
K.E.\ McGowan, S.\ Zane, M.\ Cropper, J.A.\ Kennea, F.A.\ C\'ordova,
C.\ Ho, T.\ Sasseen, and W.T.\ Vestrand. 2004.
Astrophys.\ J.\ 600 (2004) 343.

\bibitem{zp04}
V.E.\ Zavlin and G.G.\ Pavlov,
Mem.\  Soc.\ Astron.\ Ital.\ (2004) in press
[astro-ph/0312326].

\bibitem{pz03}
G.G.\ Pavlov and V.E.\ Zavlin VE,
In:  Texas in Tuscany. XXI Texas Symposium on Relativistic Astrophysics,
R.\ Bandiera, R.\ Maiolino, and F.\ Mannucci (eds), 319 (Singapore:
World Scientific Publishing) 2003.

\bibitem{motchetal03}
C.\ Motch, V.E.\ Zavlin, and F.\ Haberl,
Astron.\ Astrophys. 408 (2003) 323.

\bibitem{gyp01}
O.Y.\ Gnedin, D.G.\ Yakovlev, and A.Y.\ Potekhin,
MNRAS 324 (2001) 725.

\bibitem{pycg03}
A.Y.~Potekhin, D.G.~Yakovlev, G.~Chabrier, and O.Y.~Gnedin,
Astrophys.\ J. 594 (2003) 404.

\bibitem{pal88}
M.\ Prakash, T.L.\ Ainsworth, and J.M.\ Lattimer,
Phys.\ Rev.\ Lett.\  61 (1988) 2518.

\bibitem{ls01}
U.\ Lombardo and H.-J.\ Schulze,
In: Physics of Neutron Star Interiors,
D.\ Blaschke, N.K.\ Glendenning, and A.\ Sedrakian (eds.),
30 (Springer: Berlin) 2001.

\bibitem{frs76}
E.G.\ Flowers, M.\ Ruderman, and P.G.\ Sutherland, 
Astrophys.\ J.\ 205 (1976) 541.


\bibitem{cb03}
P.\ Chang and L.\ Bildsten,
Astrophys.\ J.\  585 (2003) 464.

\bibitem{kyg02}
A.D.\ Kaminker, D.G.\ Yakovlev, and O.Y.\ Gnedin,
Astron.\ Astrophys.\  383 (2002) 1076.

\bibitem{lpph91}
J.M.\ Lattimer, C.J.\ Pethick, M.\ Prakash, and
P.\ Haensel,
Phys.\ Rev.\ Lett.\  66 (1991) 2701.

\bibitem{pethick92}
C.J.\ Pethick,
Rev.\ Mod.\ Phys.\  64 (1992) 1133.

\bibitem{yh03}
D.G.\ Yakovlev and P.\ Haensel,
Astron.\ Astrophys.\  407 (2003) 259.

\bibitem{khy01}
A.D.\ Kaminker, P.\ Haensel, and D.G.\ Yakovlev,
Astron.\ Astrophys.\  373 (2001) L17.

\bibitem{dh02}
F.\ Douchin and P.\ Haensel,
Astron.\ Astrophys.\ 380 (2001) 151.

\bibitem{gus04}
M.E.\ Gusakov, A.D.\ Kaminker, D.G.\ Yakovlev, and O.Y.\ Gnedin,
Astron.\ Lett.\ 30 (2004) 758 


\end{thebibliography}
\end{document}